\documentclass[12pt,preprint]{aastex}
\usepackage{psfig}

\shorttitle{Filaments in M87}
\shortauthors{Sparks et al.}

\begin{document}


\title{X--ray and Optical Filaments in M87}


\author{William B. Sparks, Megan Donahue}
\affil{Space Telescope Science Institute, 3700 San Martin Drive,
    Baltimore, MD 21218}

\and

\author{Andr\'es Jord\'an\footnote{Claudio Anguita Fellow}, Laura Ferrarese, Patrick C\^ot\'e}
\affil{Dept. of Physics \&\ Astronomy, Rutgers University,
136 Frelinghuysen Road,
Piscataway, NJ 08854.}



\begin{abstract}
We compare a very deep X--ray image of M87, at the center of the Virgo Cluster,
to high-quality optical images of the low excitation emission-line gas in the
same region. There are striking coincidences of detail between the two.
We explore the possiblity that this represents a
thermal interaction between hot gas at $10^7$~K and warm gas at $10^4$~K.
We find two temperatures are present in the X--ray gas, with the lower more prevelant in
the vicinity of the optical filaments.
Electron conduction from the hot phase to the cooler one
provides a quantitatively acceptable
energy source for the optical filaments, and we show additionally that it
can do so for the brightest X--ray cluster, Perseus.
If operative, conduction in the presence of gas-rich galaxy mergers,
may explain the
presence of ``cool cores'' in clusters of galaxies.
\end{abstract}




\section{Introduction}

Recent observations with the new generation of X--ray satellites, XMM-Newton
and Chandra, have shown that the conventional cooling-flow
theory (Fabian 1994) is no longer tenable.
Data from XMM-Newton has revealed that while there is relatively cool X--ray
emitting gas in the centers of galaxy clusters, there is an absence of predicted
cooler gas below $\sim 10^7$~K
(Molendi \& Pizzolato 2001,
B\"ohringer et~al.\ 2002, Matsushita et~al.\ 2002, Molendi 2002, Donahue \& Voit 2003).
Two mechanisms in particular have been invoked to compensate for
energy lost through X--ray radiation, namely heating by an intermittently
active nucleus
(e.g.\ Tucker \& Rosner 1983, Reynolds et~al.\ 2001, Ruszkowski \& Begelman 2002, Kaiser \& Binney 2003)
and thermal electron conduction of heat into the inner regions from
the vast reservoir of thermal energy in the cluster
e.g.\ Tucker \& Rosner 1983, Rosner \& Tucker 1989, Sparks, Macchetto, \& Golombek 1989, Sparks 1992,
Voigt et~al.\ 2002, Fabian, Voigt, \& Morris 2002, Zakamska \& Narayan 2003).

The Virgo cluster of galaxies is the nearest cluster with substantial X--ray
emission, hosting a small but typical cool core
and the well-studied active radio galaxy M87.
XMM-Newton spectroscopy of the core of the Virgo cluster was amongst the pivotal
work that showed large amounts of cooling gas are not present.
The lower limit to the temperature found by XMM-Newton was 0.8~keV, or $0.9\times 10^7$~K (Molendi 2002).
The proximity of M87, 16.1~Mpc (Tonry 2001), makes it an attractive target for detailed
physical studies, with spatial resolution of 1~arcsec corresponding to 77~pc.
Here, we describe a very deep Chandra X--ray image and show that it displays
a wealth of fine structure on the arcsec scale. Comparison with optical
images reveals that these filaments are closely related to the $H\alpha+[NII]$ emitting
warm gas at $T\sim 10^4$~K.
This in turn suggests interaction between two very different gas phases,
and hence provides clues to the physical transport processes operating in the cluster center.

From an optical perspective, the emission filament system was studied
by Sparks, Ford, \& Kinney (1993), and earlier, by Ford \& Butcher (1979).
Sparks et~al.\ (1993) found that the emission filaments are dusty, that the innermost ones
are likely to be outflowing due to interaction with the jet, but that the outer
filaments are infalling. The overall excitation state is low, typical of
filament systems in elliptical galaxies and galaxy clusters.
The kinematics and geometry inferred from the dustiness and correlation with
radio source structures led Sparks et~al.\ (1993) to propose that the filaments have
their origin in a small merger event.
Weil, Bland-Hawthorn \& Malin (1997) found evidence of diffuse low surface brightness light in M87
at large radii to the Southeast, which they interpreted as tidal debris of a recent encounter.

Young, Wilson, \& Mundell (2002) presented an analysis of a 40~ks Chandra X--ray
image of the Virgo cluster. They showed that the central regions are cooler
than the outer parts of the cluster, and that there are localized cool
regions associated with the complex radio source of M87, with its jet, inner lobes,
intermediate lobes, and outer, apparently buoyant, relic radio structure.
Young et~al.\ (2002) further showed that the X--ray structures to the North and East
of the nucleus correlate with optical $H\alpha+[NII]$ filaments, and
that the X--ray gas was coolest in the vicinity of the optical filaments.

Here we present a $\sim 150$~ks Chandra X--ray image, $\sim 4\times$ longer integration than
the Young et~al.\ (2002) data.
With this improved $S/N$, we show a much clearer view of the X--ray
filaments and the detail within them, and hence reveal with greater clarity
the nature of the correlation between the optical and X--ray emission.
We also utilize the spectral information of the new data to probe the
physical conditions within the filaments compared to their surroundings.
Motivated by an earlier suggestion that electron conduction from the hot gas
into the cold gas of an accretion or merger event can explain the peculiar symptoms
of a ``cooling flow'' (Sparks et~al.\ 1989), we assess whether
this mechanism could provide the energy to ionize the filament system.

\section{Observations}

\subsection{Chandra X--ray observations}

To create an X--ray image with high signal-to-noise
we merged the three longest Chandra ACIS-S observations of the Virgo cluster.
The relative astrometry of the observations, based on inspection of
point sources detected in all three observations, 
showed that no shifts or rotations were required to stack the data.  
Level-1 CTI-corrected event files were generated by running 
{\it acis\_process\_events} (CIAO 2.3). Level-2 event files were created 
by selecting events with grades 0 2 3 4 6, and status of 0. The ``good time
intervals'' supplied by the original pipeline were used. We then
examined a light curve from the S3 chip and 
omitted data obtained during observation periods affected by major flares.
We then merged all three independent observations, and
a final event file was created from a subset of the merged file
by retaining only the 0.3--3.0~keV events from the S3 back-illuminated
CCD. An exposure map 
for the merged observation was created assuming a monochromatic 
incident spectrum of 1.5~keV. The resulting map has units
photons s$^{-1}$ cm$^{-2}$ pixel$^{-1}$ and pixel size
$0.492''\times 0.492''$. A summary of the observations is reported in Table~1.
The total useful exposure time of the final image is $\approx 154$~ks.

\subsection{Narrow-band Optical Imaging}

For morphological comparison, we use the ESO/NTT 3000~sec $H\alpha+[NII]$
image of Macchetto et~al.\ (1996). The image was constructed from two separate
exposures of 1500~sec each obtained on the ESO 3.4-m NTT,
and processed in the conventional fashion as described
in Macchetto et~al.\ (1996).
Continuum subtraction was performed using a long $R-$band exposure.
This long exposure was saturated in the central regions, and displayed
a low-level ghost weakly visible in Fig.~1.
These qualities do not affect a morphological comparison. To carry out
a quantitative flux calibration, we used the lower
signal-to-noise, but well-calibrated image
from KPNO presented in Sparks et~al.\ (1993).
That image was constructed from $5\times $ 120~sec exposures and
the reduction procedures and flux calibration are documented 
in Sparks et~al.\ (1993).

\section{Morphology and Morphological Comparisons}

Fig.~1 shows the new, high $S/N$ X--ray image. The image has been adaptively smoothed
to retain a $S/N=10$ per pixel, and is displayed on a logarithmic scale. There is a wealth of detail
and filamentary structure as well as arcs, edges and voids within the hot gas.

For comparison to the optical line-emitting gas, we resampled the optical
and X--ray images to be spatially registered and blinked between them.
Fig.~2 shows the X--ray image next to the $H\alpha+[NII]$ image, and labels
some of the features common to both (given in boldface in the text).
Beginning in the Southeast corner, the knotty complex of $H\alpha+[NII]$ emission
coincides with the outer edge of the Southeast inner radio lobe (Sparks et~al.\ 1993).
The X--ray emission shows a similar complex, also brightest on the edge closest
to the nucleus. Just North of this complex, where
optical filaments appear to cross, we see both brighter optical
emission and a knot in the X--ray emission.
To the East of the knot the optical emission diverges from this point in a double
strand, and fades, as also does the X--ray emission.
From the knot, the $H\alpha+[NII]$ filament runs almost straight North-West
and then kinks to run East-West. The same behavior is present in the X--ray image.
Moving to the East, the filaments brighten in a bar as they converge with an arc
located to the East of the nucleus. This arc of emission
ends at a bright knot South of the nucleus. North of the nucleus and emerging from it,
there is a very bright loop of emission, also in both images, reminiscent of a Solar coronal loop.

There are a few places where the maps differ: to the Southwest of the nucleus,
there are X--ray knots without optical counterparts, while the
$H\alpha+[NII]$ filaments due West of the tip of the jet have little
or no X--ray counterpart.
The bright optical filament near the jet running Northwest does not
have a strong X--ray counterpart.
Also, the flux ratios are not identical even where the morphology is similar.
Overall, however, there are clear coincidences in shape and distribution
between the two datasets.

This is a similar situation to that found in the Perseus Cluster
by Fabian et~al.\ (2003) where optical filaments also have unambiguous
counterparts in a very deep X--ray image.

\section{Discussion}

\subsection{Physical properties inferred from X--ray data}

In order to compare physical conditions in the filaments and inter-filament regions,
we extracted X--ray events from 6~apertures from the longest single
observation of 98,664 seconds (OBSID 2707).
Fig.~4 shows the apertures used, each having
between 2,000 and 10,000 X--ray counts total, and $>25$~counts per spectral element.
B1 and B2 are box shaped and lie on
bright X-ray features with counterparts in the $H\alpha$ image. C1--4 are
circular and lie at nearly the same distance from the nucleus, but are not on bright
filamentary features. A small correction was made for background contamination
and weighted RMF and ARF files were created for each aperture.
XSPEC 11.2.0 was used to analyze the resulting spectra. The HI column
was kept fixed at $1.8 \times 10^{20}$ cm$^{-2}$ (Lieu et~al.\ 1996).
The spectral fits are summarized in Table~2.
All required two temperature components, and
we fitted a two-temperature MEKAL model with a single metallicity for both
components. Densities were computed assuming pressure  equilibrium.
Quoted uncertainties are 90\% confidence intervals based on propagated
statistical errors only. The emission was assumed to arise from a spherical
volume defined by the circular apertures and from a cylindrical volume defined by the
elongated box apertures, and we neglected projection through the cluster (Molendi 2002).
To provide a visual sense of the X--ray spectral data, we also constructed a
red/green/blue image using X--ray passbands from 0.5 to 1~kev (red), 1 to 2~keV (green) 
and above 3~keV (blue), shown in Fig.~5. 
The filaments are more pronounced in the lower energy bands, revealing this
with their redder color in Fig.~5 and indicative of a lower mean temperature
within the filaments.

The detailed spectral fits find two temperature components within the X--ray plasma,
0.8 and 1.6~keV, consistent with earlier findings of Young et~al.\ (2002), Molendi (2002),
for {\it both\/} the filament and inter-filament regions.
Interestingly, the temperature values for the two regions are the same
within the uncertainties. What differs, though, is the relative proportion of cool and hot gas.
In the $H\alpha$ filaments, the cool-gas filling factor is of order twice that of the inter-filament region.
The pressures derived from the X--ray fitting, 0.6---$2.5\times 10^{-9}$~dynes~cm$^{-2}$, are very 
comparable to those derived from optical observations (Heckman et al. 1989), $0.6$ --- $2.4\times 
10^{-9}$~dynes~cm$^{-2}$.

\subsection{Conduction as an energy source for the optical filaments}

Sparks et~al.\ (1989), primarily based on optical observations of the dusty emission filaments in the
Centaurus Cluster of galaxies (NGC~4696),
proposed that cool, merging gas falling into a hot X--ray corona
could be energized by thermal conduction from the hot phase.
This would cause optical line emission from the cooler gas
and locally enhance the X--ray emission. The two phases should
consequently display spatial similarities.
Cowie \& McKee (1977) pointed out that if temperature gradients are sufficiently
high, classical electron conduction (Spitzer 1962) overestimates the actual energy flux. The heat conductivity
``saturates'' and is limited by the electrons available to cross a surface.
This could arise for example if the system is in an unrelaxed dynamic, transient state, or if
length scales are smaller than a critical value, related to the electron mean free path.

To see if conduction can provide a plausible energy source for the optical line emission,
we note that the fraction of energy emitted
as $H\alpha+[NII]$ line radiation is expected to be of order 2---6\% of the total optical line cooling 
(Voit et~al.\ 1994). Allowing for energy losses from the interface radiation and to infrared
emission of heated dust, we assume $\sim 1$\%\ of the available heat energy re-emerges as $H\alpha+[NII]$.
The topology of the filament system is unknown, however the surface area is at least twice the
projected surface area (front and back) and is potentially much higher, e.g.\ if
the filaments are ``clouds'' of small blobs, so conservatively we assume
the actual area is $3\times$ the apparent. The saturated heat conductive flux per unit area is
\begin{equation}
q_{sat} = 0.4 \left( {2 k T_e\over \pi m_e}\right)^{1/2}n_e k T_e
\approx  0.0542 T_7^{3/2} n_{0.1}
\hbox{ erg s$^{-1}$ cm$^{-2}$ }
\end{equation}
where $T_7 = T_e/10^7$K and $n_e$ is the electron density, with $n_{0.1} = n_e/0.1$~cm$^{-3}$.
The classical Spizter heat flux per unit area is
\begin{equation}
q = \kappa \nabla T
\approx 0.55 T_7^{7/2}/L_{pc}
\hbox{ erg s$^{-1}$ cm$^{-2}$ }
\end{equation}

For a temperature $T \approx 10^7$~K and an electron density $\approx 0.1$~cm$^{-3}$,
the conductivity saturates if the length scale is $\sim 10$~pc.
In M87, the X--ray filaments and optical filaments appear marginally resolved,
although since we may be viewing sheets or ribbons, the interface length scale could be smaller.
For reference, we adopt a length scale of $L \sim 100$~pc at 20~arcsec radius,
and assume it is inversely proportional to radius.
Two temperatures are needed to describe the X--ray gas in M87, see above and Molendi (2002).
The lower has $T\approx 0.7$---$0.9$~keV and the higher $\approx 1.4$---$1.6$~keV,
and since the $H\alpha+[NII]$ filaments are associated with a higher proportion
of cooler gas (\S~4.1, and Young et~al.\ 2002),
we adopt a temperature of 0.9~keV, and extrapolate the Northeast density profile of
Young et~al.\ (2002) inwards for the simplest model temperature/density profile ``YWM.''
Kaiser (2003) develops a self-consistent entropy-based model of the hot gas in M87,
which may also be transformed to a temperature/density profile.
Fig.~2(a) plots the observed $H\alpha+[NII]$ flux versus radius, 
with the YWM and Kaiser models in the saturated regime, and
in the classical regime with the same reference length scale as above.
The figure shows that saturated electron conduction can easily provide enough energy to
supply the optical filaments, and that classical unsaturated conduction can too, if it is able to operate
at essentially the Spitzer level for our assumed length scale.
The saturated regime would be entered if the actual length scales are
small ($\sim 10$~pc). The classical domain would be valid for larger length scales,
and the effectiveness of both will depend on the topology of the magnetic fields threading the system.

Fabian et~al.\ (2003) showed that similar optical/X--ray coincidences
occur in the filaments of the Perseus Cluster (NGC1275).
The X--ray filaments are $<\sim 500$~pc in width, the inner temperature is
$\approx 3$---$4\times 10^7$~K, and the density in the central region is $\sim 0.05$~cm$^{-3}$.
For these values, the electron conductivity is entering the saturated regime.
In the absence of a definitive temperature/density profile for Perseus, we adopt
two representative models, one optical and one X--ray based.
Heckman et~al. (1989) present a pressure profile for the center of the Perseus
cluster derived from optical emission line ratios.
Assuming the optical gas is in pressure equilibrium with
the X--ray gas, and that the X--ray gas temperature is constant at
$T\approx 4\times 10^7$~K, a reasonable description of their pressure profile is
$$P \approx 10^{-9}(1.36 \hbox{kpc}/R)^{0.6}\hbox{dyne.cm}^{-2}$$
Fabian et~al. (2000, 2003) quote a steeper density dependence, based on X--ray data,
$n_e \propto 1/R$. Hence we also include a fiducial model with that behavior, normalized to
have density 0.04~cm$^{-3}$ at a radius 1~arcmin and $T = 3\times 10^7$~K.
Fig.~2(b) overlays saturated conductive models
based on these temperature/density profiles on the optical emission filament
surface brightness (Fig.~7 of Conselice et~al.\ 2001) for Perseus.
Here, we see that the saturated electron conduction predictions of these
representative models, {\it derived from the properties of the coronal plasma only\/}, are remarkably 
close to the actual {\it optical data\/}.  We conclude that, given the uncertainties, saturated electron 
conduction could be providing the power to the optical filament systems.

If the energy loss from the (optical/UV/IR) filament system does
represent the heat being drained from the coronal phase, then 
it is energetically significant, overwhelming the losses due to X--rays at least locally.
The energy flow rather than the total energy content dominates in
this scenario, and radiation governs the ultimate loss of energy,
primarily in the optical and infrared regions.
The consequent cooling of the cluster core will enhance its X--ray
emissivity as the density increases to maintain pressure equilibrium.
The essential features of a ``cool core cluster'' (or cooling-flow cluster)
are hence induced.

There are of course caveats to this interpretation.  Topology,
projection effects, magnetic fields and details of the interface impose
large uncertainties on the available energy.  If the cool gas is
significantly in the foreground or background, then the available
energy can be much less.  If magnetic fields shield the filaments, the
heat may not be able to flow into the filaments.  Evaporating and
condensing clouds in the presence of radiation and conduction have been
studied by McKee \& Cowie (1977) and Begelman \& McKee (1990), (though
we note that McKee \& Cowie 1977 eliminate the important thermally stable
$10^4$~K phase from their analysis).  In reality, the details of
timescales, dynamical motions and inner and outer boundary conditions
will all enter into a correct description of the physics of these
complex regions, and it may well be that other transport processes
are equally, or more, effective than electron conduction.
We accept that there may indeed be other viable interpretations of the data.
If indeed, it can be shown that the optical filaments are powered by a mechanism other
than energy flow from the hot plasma, then we may infer that conduction
would have to be suppressed since it represents a significant energy source.
However, to counterbalance this, we note that the models presented above,
derived from properties of the {\it coronal X--ray gas\/},
give acceptable agreement to the {\it optical\/} $H\alpha$ data, without adjustment of free parameters.

Fabian et~al.\ (2003) draw attention to the similarity of the Perseus filaments to a rising spherical cap bubble,
which has an analogue in M87: the Eastern low-frequency (intermediate) radio lobe
is interacting with the halo X--ray emitting gas, and drawing it out from
the center (Churazov et~al.\ 2001).
Sparks et~al.\ (1993) showed evidence (from the velocity field plus absorption due
to dust) that the innermost optical filaments of M87 are flowing out in the vicinity of the jet.
Hence it may be that in such systems, the large and rapid energy release associated
with the triggering of an AGN is providing, in a general sense, heating and an outflow
in the central regions of the clusters, and that in both Virgo and Perseus the optical
and X--ray filaments towards the center are expanding away from the nucleus, and not
infalling. The outer filaments, by contrast, do appear to be infalling, and it is
reasonable to speculate that the AGN activity was triggered by the same infall or merger
event that is interacting with the X--ray halo.
Since we are dealing in concept with fundamental physical and transport processes,
if valid here, then this conclusion must have wide validity.

Given the dynamic and violent nature of cluster centers---regions that are influenced by 
mergers, active nuclei and energy transport processes---we suggest that the cool-core
phenomenon is transient and merger induced.
It is unlikely we are witnessing a steady-state even in the X--rays.
If merger events have triggered nuclear activity,
and if strong thermal interactions arise between hot and cold gas,
then a logical consequence is that
prominent X--ray peaks of cool-core galaxy clusters are themselves
transient and only intermittently present as 
the stochastic evolution and assembly of galaxy clusters proceeds.

\section{Conclusions}

Our primary goal has been to present a very deep X--ray image of the center of the Virgo
cluster, and to stress an important empirical observation,
namely that warm gas with $T \sim 10^4$~K emitting strong optical line
emission is strikingly similar to a phase which is a thousand times
hotter and radiates at X--ray wavelengths. This correlation may be offering important clues
to the nature of transport processes in the cluster interstellar medium,
since it implies that two dramatically distinct phases, close in space,
``know about one another'' and are in some form of thermal communication.
X--ray spectral analysis shows that the hot phase itself has two temperature
components, and that the lower of these is more prevelant
in the vicinity of the optical filaments.

We build on these observations to show, quantitatively, that
electron conduction (possibly saturated) can provide the
energy seen in the form of emission line radiation in these and other
optical filaments. We do not claim that this is the {\it only\/} viable model.
There are important uncertainties particularly with regard to magnetic field
structures and strengths, and conduction would have to proceed at or close to 
Spitzer or saturated levels at least in one direction. Nevertheless, the model appears promising
and offers an acceptable {\it quantitative\/} relationship between $10^7$~K coronal
and $10^4$~K plasma properties. If the material of the filaments has an external merger origin,
then other ancillary properties such as their dustiness have a natural explanation.

Given the dynamic and violent nature of galaxy cluster centers,
influenced by mergers, active nuclei and energy transport processes,
we suggest that the cool-core phenomenon is transient and merger induced.



\acknowledgments

The Space Telescope Science Institute is
operated by the Association of Universities for Research in Astronomy,
Inc., under NASA contract NAS5-26555.
Support for A.~Jord\'an was provided by the National Science Foundation
through a grant from the Association of Universities for Research
in Astronomy, Inc., under NSF cooperative agreement AST-9613615,
and by Fundaci\'on Andes under project No.C-13442.
P. Cot\'e gratefully acknowledges support for this research provided by NASA
LTSA grant NAG5-11714 and funding for Chandra program CXC03400562.

\clearpage 

\begin{figure}
\plotone{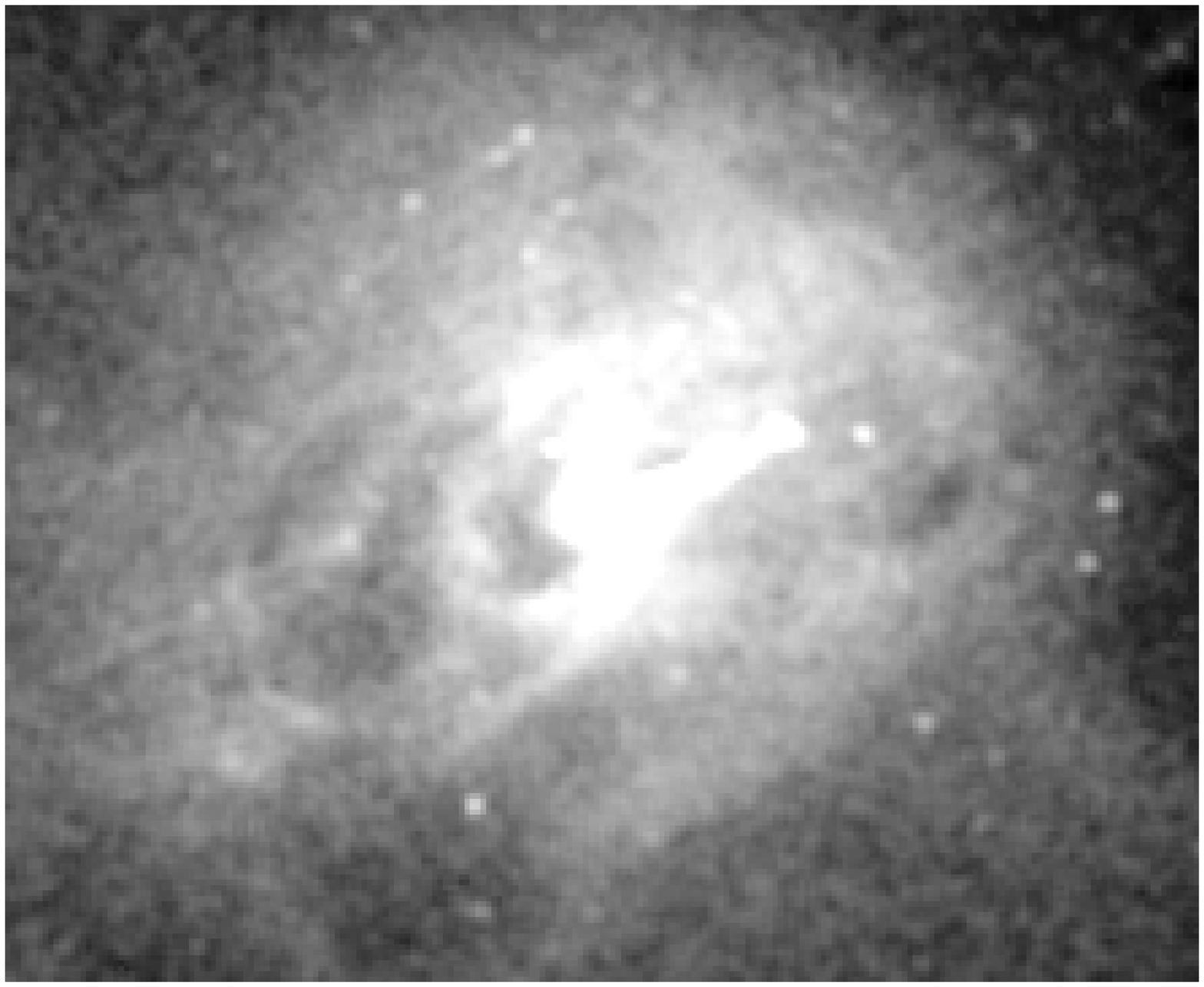}
\caption{Adaptively filtered soft X--ray emission, the image is $1.8\times 1.5$ arcmin, North up and East to the left.
\label{fig1}}
\end{figure}

\clearpage 

\begin{figure}
\plotone{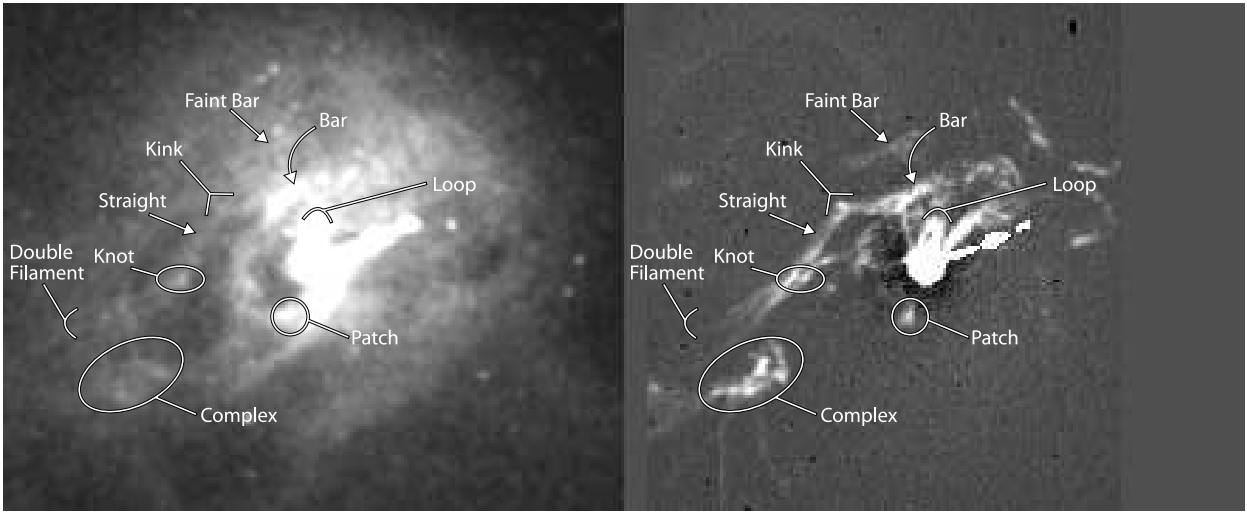}
\caption{Adaptively smoothed soft X--ray emission, 154~ksec of Chandra ACIS-S observations (left), compared
to the optical $H\alpha + [NII]$ filaments system (right). The images are both $1.8\times 1.5$ arcmin.
\label{fig2}}
\end{figure}

\clearpage 

\begin{figure}
\epsscale{1.0}
\plotone{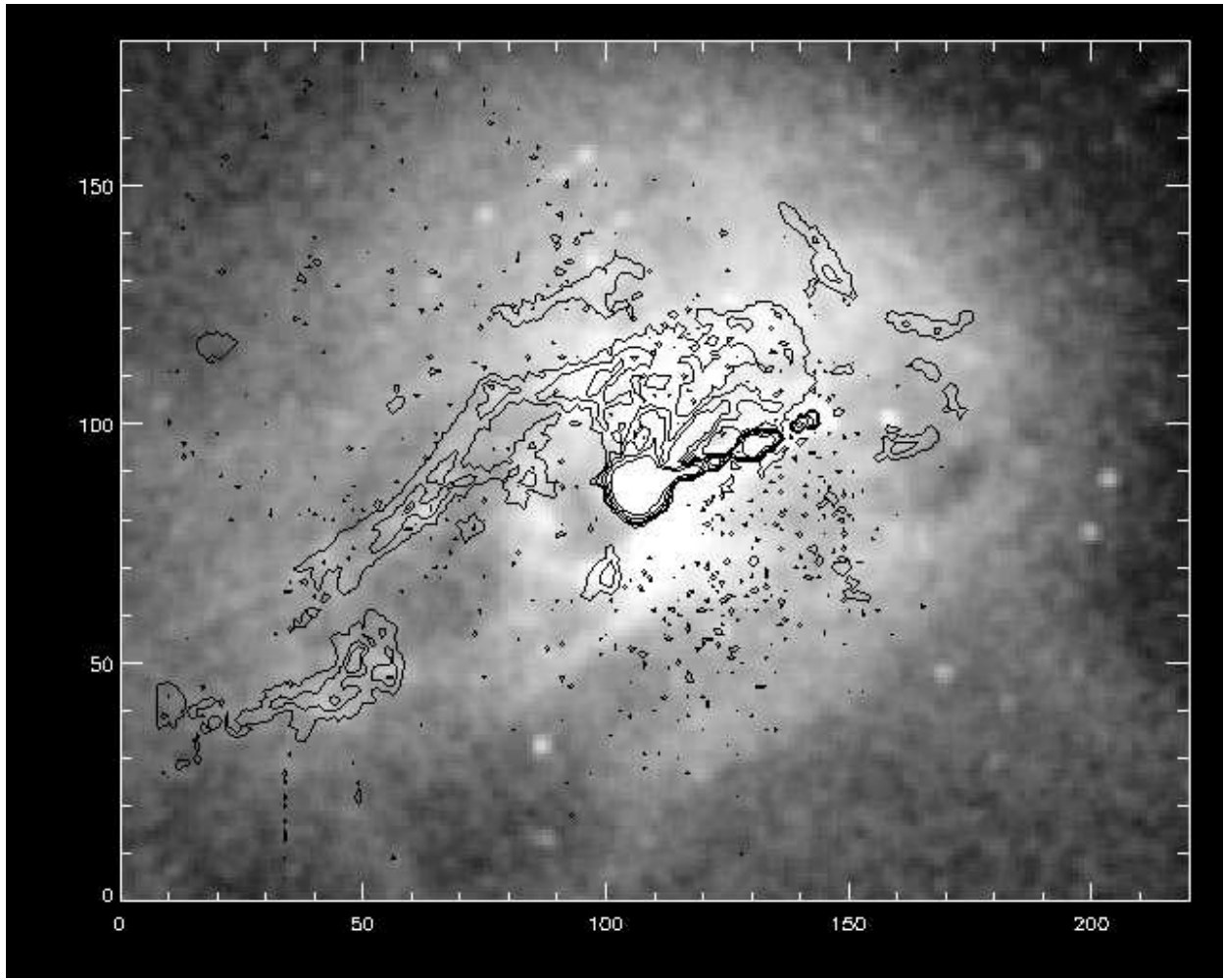}
\caption{Overlay of $H\alpha$ contours on Chandra X--ray image using the images of Figs.~1 and 2.
\label{fig3}}
\end{figure}

\clearpage

\begin{figure}
\epsscale{1.0}
\plotone{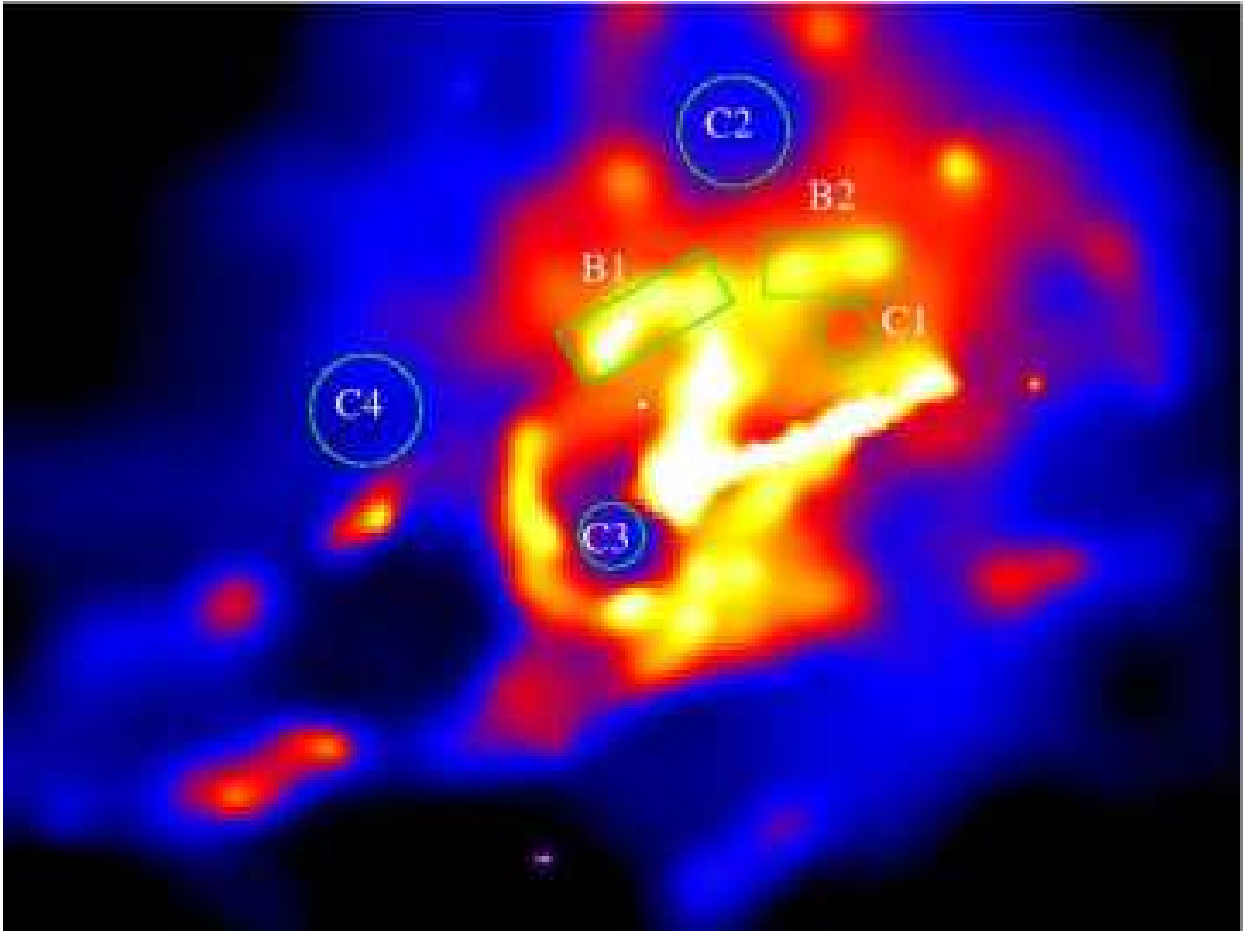}
\caption{False color image of the X--ray emission of M87 showing apertures used in X--ray spectral analysis, as described in the text.
\label{fig4}}
\end{figure}

\clearpage

\begin{figure}
\epsscale{1.0}
\plotone{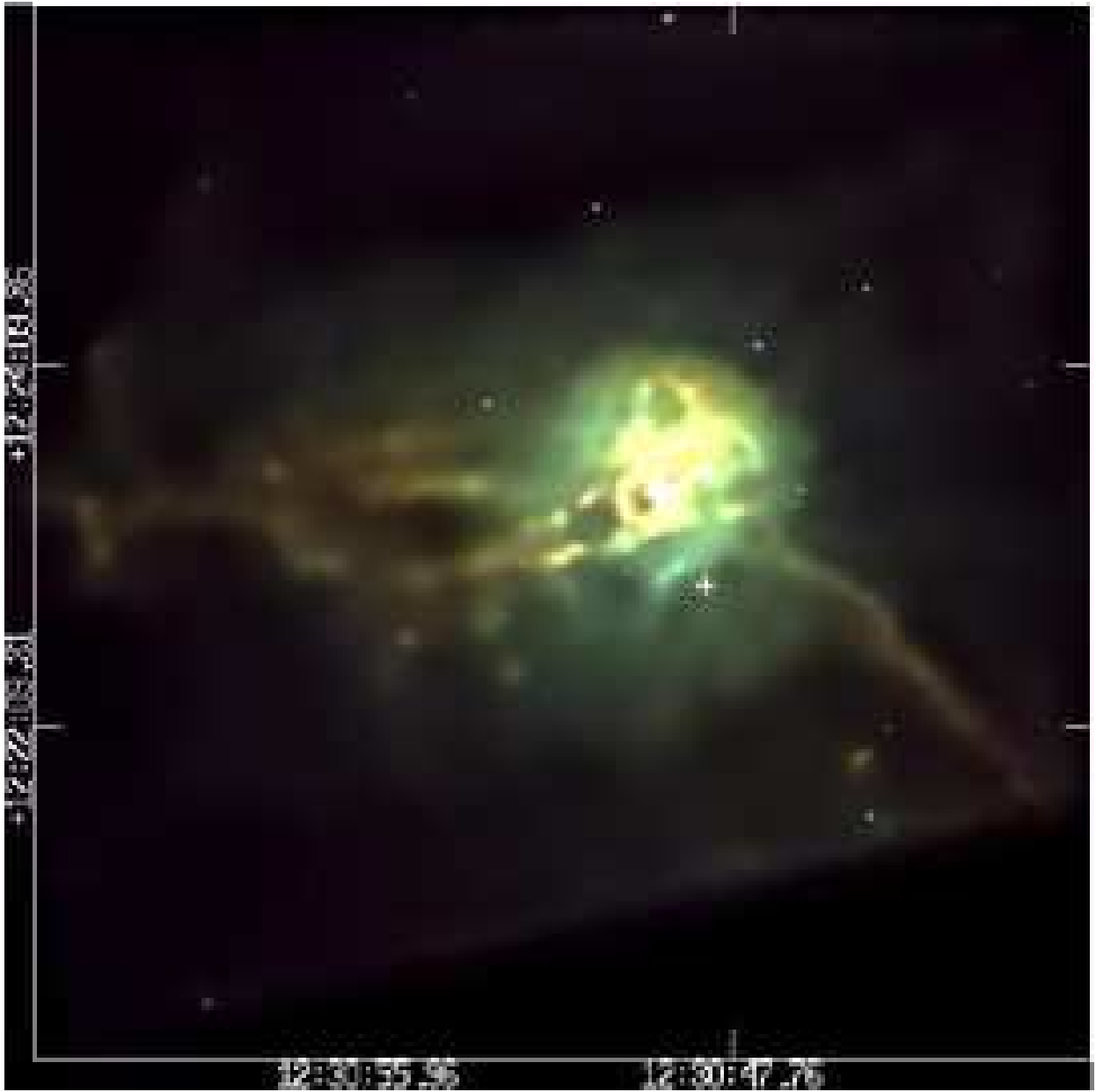}
\caption{X--ray color image derived using the spectral information of the Chandra data. Redder regions, typically the regions where the filaments lie, have a more dominant cool component in the two-temperature hot gas.
\label{fig5}}
\end{figure}

\clearpage

\begin{figure}
\centering
\includegraphics[height=2.5in]{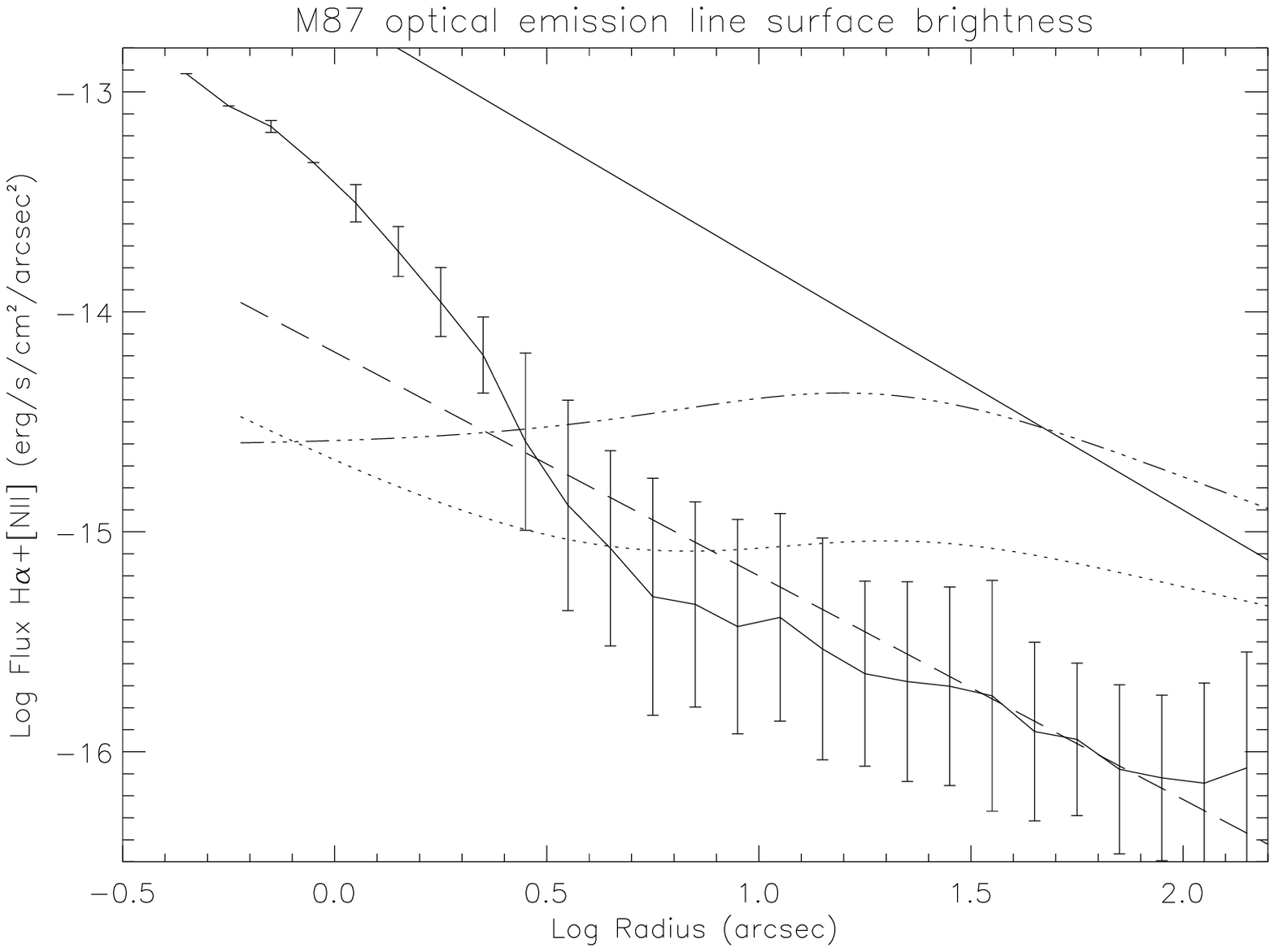}\includegraphics[height=2.5in]{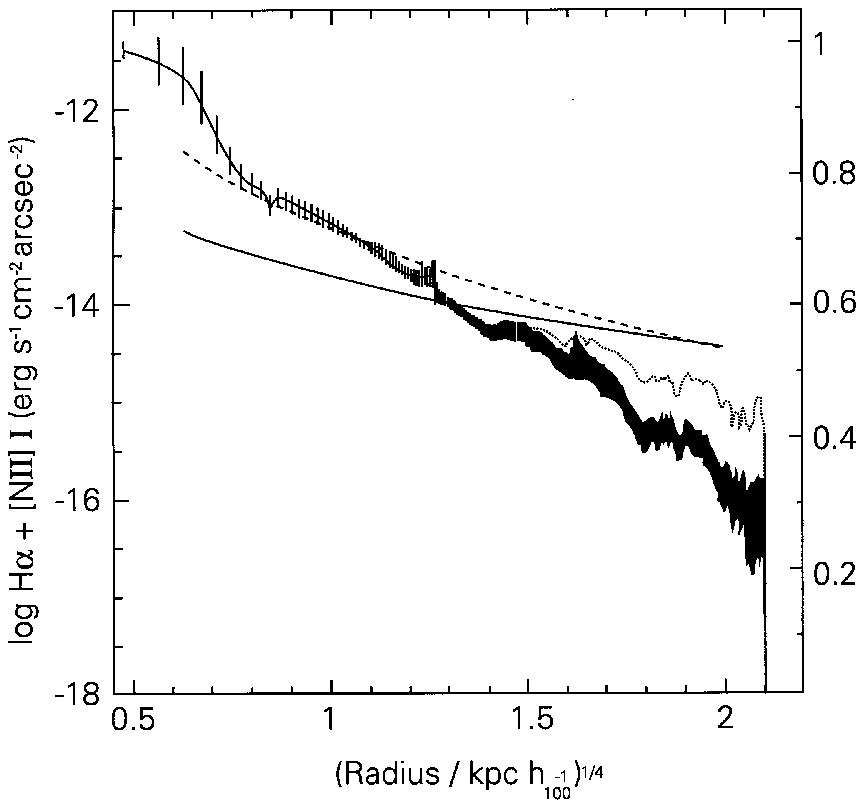}
\caption{(Left) The optical $H\alpha+[NII]$ surface brightness profile of M87 derived from Sparks, Ford \& Kinney (1993). For comparison, the energy from electron conduction assuming 1\% is re-radiated in these lines. Solid is YWM saturated; long dashes YWM  classical for a length scale $100(20/r$~arcsec)~pc;
dot-dash is Kaiser saturated, dotted is Kaiser classical.
(right)~Two saturated conduction models for Perseus, overlaid on the $H\alpha$ surface brightness from Fig.~7 of Conselice et~al.\ 2002. The conduction models (running in shallow curves across the center of the diagram)
are derived from inferred properties of the coronal gas and are compared to the optical observations of 
$10^4$~K gas. The models have density $\propto R^{-0.6}$ (solid, lower) and $\propto R^{-1}$ (dashed, upper) curves.
\label{fig6}}
\end{figure}







\clearpage
\centering
\begin{deluxetable}{ccccc}
\tablewidth{5.5in}
\tablecolumns{5}

\tablecaption{Chandra Observations of M87}

\tablehead{
\colhead{Seq Num} & \colhead{Obs ID} & \colhead{Date} & \colhead{Original Exposure} & \colhead{Useful Exposure} \\
        &       &        &    \colhead{(seconds)}     &       \colhead{(seconds)}
}
\startdata
700024  &  0352 &  July 29, 2000 & \phn40,000 & 37,112 \\ 
400187  &  2707 &  July  \phn5, 2002 & \phn23,000 & 18,072 \\
400187  &  3717 &  July  \phn6, 2002 & 105,000 & 98,664 \\
\enddata
\end{deluxetable}

\begin{deluxetable}{cccccccccc}
\tabletypesize{\footnotesize}
\tablecolumns{7} 
\tablewidth{0pc} 

\tablecaption{X--ray spectra of selected regions} 
\tablehead{ 
\colhead{Aperture} & \colhead{$n_e(\hbox{cool})$}   & \colhead{$kT_c$} &
\colhead{$n_e(\hbox{hot})$} & 
\colhead{$kT_h$}    & \colhead{$P$}   & \colhead{$f_c$}    &
\\

\colhead{name} & \colhead{cm$^{-3}$}   & \colhead{keV} &
\colhead{cm$^{-3}$} & 
\colhead{keV}    & \colhead{$10^{-9}$~dynes~cm$^{-2}$}   & \colhead{}

}
\startdata 
B1         & $1.18\pm0.08$ & 0.69 (0.65--0.74) & $0.56\pm0.06$ & 1.45
(1.22--1.50) & $2.5\pm0.3$ & $0.07\pm0.03$ \\
B2         & $0.92\pm0.05$ & 0.82 (0.78--0.87) & $0.49\pm0.04$ & 1.54
(1.42--1.72) & $2.3\pm0.2$ & $0.11\pm0.04$ \\
C1         & $0.51\pm0.05$ & 0.77 (0.71--0.83) & $0.28\pm0.05$ & 1.4\phn
(1.26--1.56) & $1.2\pm0.2$ & $0.11\pm0.05$ \\
C2         & $0.25\pm0.06$ & 0.78 (0.63--1.22) & $0.13\pm0.03$ & 1.5\phn
(1.4--1.7)\phn\phn    & $0.6\pm0.4$ & $0.05\pm0.04$ \\
C3         & $0.64\pm0.10$ & 0.80 (0.74--0.94) & $0.32\pm0.06$ & 1.6\phn
(1.4--2.0)\phn\phn   & $1.6\pm0.3$ & $0.04\pm0.02$ \\
C4         & $0.35\pm0.05$ & 0.87 (0.74--1.02) & $0.17\pm0.03$ & 1.8\phn
(1.6--2.2)\phn\phn   & $0.9\pm0.3$ & $0.03\pm0.01$ \\
\enddata 
\end{deluxetable}

\end{document}